\documentclass[11pt]{article}
\title
{Construction of a Non-2-colorable $k$-uniform Hypergraph with Few Edges}
 \author{Heidi Gebauer
 \thanks{Institute of
 Theoretical Computer Science, ETH Zurich, CH-8092 Switzerland. Email:
 gebauerh@inf.ethz.ch. } 
 }

\usepackage{amsmath, amsfonts}
\usepackage{amsthm}
\usepackage{subfigure}
\usepackage[dvips]{graphicx}

\begin{document}
\bibliographystyle{plain}
\maketitle
% Theorems, etc.
\newtheorem{theo}{Theorem} [section]
\newtheorem{defi}[theo]{Definition}
\newtheorem{lemm}[theo]{Lemma}                                                                                                                                                                                                                                                                                                                                                                                                                                                                                                                   
\newtheorem{obse}[theo]{Observation}
\newtheorem{prop}[theo]{Proposition}
\newtheorem{coro}[theo]{Corollary}
\newtheorem{rem}[theo]{Remark}
\newtheorem{claim}[theo]{Claim}

%\newtheorem{theorem}{Theorem}
%\newtheorem{claim}{Claim}
%\newtheorem{lemma}{Lemma}
%\newtheorem{propos}{Proposition}
%\newtheorem{conjecture}{Conjecture}
%\newtheorem{problem}{Problem}
%\newtheorem{corol}{Corollary}
%\newcommand{\Proof}{\noindent{\bf Proof.}\ \ }
%\newcommand{\Remarks}{\noindent{\bf Remarks:}\ \ }
%%Probability
\newcommand{\whp}{{\bf whp}}
\newcommand{\prob}{probability}
\newcommand{\rn}{random}
\newcommand{\rv}{random variable}
%Hypergraphs
\newcommand{\hpg}{hypergraph}
\newcommand{\hpgs}{hypergraphs}
\newcommand{\subhpg}{subhypergraph}
\newcommand{\subhpgs}{subhypergraphs}
%Letters
\newcommand{\bH}{{\bf H}}
\newcommand{\cH}{{\cal H}}
\newcommand{\cT}{{\cal T}}
\newcommand{\cF}{{\cal F}}
\newcommand{\cG}{{\cal G}}
\newcommand{\cD}{{\cal D}}
\newcommand{\cC}{{\cal C}}

\newcommand{\ideg}{\mathsf {ideg}}
\newcommand{\lv}{\mathsf {lv}}
\newcommand{\nga}{n_{\text{game}}}
\newcommand{\avdaneg}{\overline{\deg}}
\newcommand{\ed}{e_{\text{double}}}

\newcommand{\danger}{\mathsf {dang}}
\newcommand{\avdanan}{\overline{\danger}}

%commands min deg c
\newcommand{\degb}{\deg_{B}}
\newcommand{\degm}{\deg_{M}}

\newcommand{\avd}{\overline{D}}
\newcommand{\pr}{\mathsf {Pr}}

\begin{abstract} 
We show how to construct a non-2-colorable $k$-uniform hypergraph with 
$(2^{1 + o(1)})^{k}$ edges. By the duality of hypergraphs and monotone CNF-formulas this gives an unsatisfiable monotone $k$-CNF with $(2^{1 + o(1)})^{k}$ clauses.
\end{abstract}

\section{Introduction}

We will show the following.
\begin{theo} \label{theo:keyclaim}
  For every $l \leq k$ we can construct a non-2-colorable $k$-uniform hypergraph with $m(k,l) = \binom{2l - 1}{l} \cdot \left(\frac{2^{l} k}{l} \right)^{l} \cdot  \dbinom{\frac{2^{l}}{l} k}{\frac{k}{l}}$  edges.
\end{theo}
\noindent
The next proposition bounds $m(k,l)$
\begin{prop} \label{prop:evalformula}
 We have $m(k,l) \leq 2^{2l + l^{2}} \cdot k^{l} \cdot 2^{k} e^{\frac{k}{l}}$. In particular, $m(k, \log k) \leq (2^{1 + o(1)})^{k}$.
\end{prop}
\noindent
Hence we obtain a non-2-colorable hypergraph with few edges.
\begin{coro} \label{coro:suitablehypergraph}
 We can constrcut a non-2-colorable hypergraph with $(2^{1 + o(1)})^{k}$ edges.
\end{coro}
\noindent
Non-2-colorable hypergraphs connect to unsatisfiable CNF formulas: For a $k$-uniform hypergraph $H$ let $H'$ denote the $k$-CNF obtained by adding for every edge 
$e = (x_{1}, x_{2}, \ldots, x_{k})$ the clauses $C_{e} := (x_{1} \vee x_{2} \vee \ldots \vee x_{k})$ and 
$C'_{e}  := (\bar{x_{1}} \vee \bar{x_{2}} \vee \ldots \vee \bar{x_{k}})$. 
Now $H'$ is monotone, i.e., every clause either contains only non-negated literals or only negated literals. Moreover, every 2-coloring $c$ of $H$ yields a satisfying assignment $\alpha$ of $H'$ (indeed, just set $\alpha(x_{i}) := 1$ if and only if $x_{i}$ is colored blue under $c$) and vice versa.
So Corollary \ref{coro:suitablehypergraph} yields the following.
\begin{coro} \label{coro:suitableCNF}
 We can construct an unsatisfiable monotone $k$-CNF with $(2^{1 + o(1)})^{k}$ clauses.
\end{coro}

\section{Constructing a Non-2-Colorable Hypergraph with Few Edges}

Throughout this section $\log$ stands for the binary logarithm. Moreover, a \emph{2-coloring} is an ordinary, not necessarily proper, 2-coloring. 

\emph{Proof of Theorem \ref{theo:keyclaim}:}
Let $k' = \frac{2^{l}}{l} k$. For every $i$, $i = 1, \ldots, 2l - 1$, we let $A_{i} := a_{i,1}, a_{i,2}, \ldots, a_{i,k'}$ be a sequence of length $k'$.
Let $c$ be a given 2-coloring. $c$ has a \emph{red majority}  (\emph{blue majority}) in the sequence $A_{i}$ if under $c$ at least $\frac{k}{2}$ elements of 
$\{a_{i,1}, a_{i,2}, \ldots, a_{i,k'}\}$ are colored red (blue). Note that $c$ has both a red majority and a blue majority in a sequence $A_{i}$ if and only if there are equally many red and blue elements. We say that $c$ has the \emph{same majority} in the sequences $A_{i_{1}}, A_{i_{2}}, \ldots, A_{i_{j}}$ if either 
$c$ has a red majority in every sequence in $\{A_{i_{1}}, A_{i_{2}}, \ldots, A_{i_{j}}\}$ or $c$ has a blue majority in every sequence in $\{A_{i_{1}}, A_{i_{2}}, \ldots, A_{i_{j}}\}$. 
\begin{prop} \label{prop:constructionforsubsetwithmajority}
For every $\{X_{1}, \ldots, X_{l}\} \subseteq \{A_{1}, A_{2}, \ldots, A_{2l-1}\}$ we can construct a $k$-uniform hypergraph $G_{X_{1}, \ldots, X_{l}}$ with at most 
$k'^{l} \dbinom{k'}{\frac{k}{l}}$ clauses such that every 2-coloring $c$ which has the same majority in $X_{1}, \ldots, X_{l}$ yields a monochromatic edge in 
$G_{X_{1}. \ldots, X_{l}}$.
\end{prop}
\noindent
Proposition directly implies Theorem \ref{theo:keyclaim}. Indeed, let $G$ be the hypergraph consisting of the union of all edges in $G_{X_{1}, \ldots, X_{l}}$ for every 
$\{X_{1}, \ldots, X_{l}\} \subseteq \{A_{1}, A_{2}, \ldots, A_{2l-1}\}$ and let $c$ be a 2-coloring of the vertices of $G$. By the pigeon hole principle, for some $X_{1}, \dots, X_{l} \subseteq \{A_{1}, A_{2}, \ldots, A_{2l-1}\}$, $c$ has the same majority for $X_{1}, \dots, X_{l}$. But then $c$ yields a monochromatic edge in $G_{X_{1}, \ldots, X_{l}}$ and so $c$ is not a proper 2-coloring of $G$.
Since $c$ was chosen arbitrarily $G$ is not properly 2-colorable. Moreover, the number of edges of $G$ is $\binom{2l - 1}{l}$ times the number of edges in 
$G_{X_{1}, \ldots, X_{l}}$, which gives the required number of edges in total. \hfill $\qed$

\emph{Proof of Proposition \ref{prop:constructionforsubsetwithmajority}:} 
Let $X_{j} = x_{j,1}, x_{j,2}, \ldots, x_{j,k'}$ for every $j$, $j = 1, \ldots, l$. 
We will now shift sequences by a certain number of elements. For every $i \in \{0, \ldots, k' - 1\}$ we let 
$X_{j}(i) = x_{j,1 + i}, x_{j, 2 + i}, \ldots, x_{j, k'}, x_{j, 1}, \ldots, x_{j, i}$.
\newline
For every $i_{1}, i_{2}, \ldots, i_{l} \in \{0, \ldots, k' - 1\}$ and for every $S \subseteq \{1, 2, \ldots, k'\}$ with $|S| = \frac{k}{l}$ we let 
$e_{i_{1}, i_{2}, \ldots, i_{l}}(S)$ denote the set of elements which are of the form $x_{j, r + i_{j}}$ with $r \in S$.
For every $i_{1}, i_{2}, \ldots, i_{l} \in \{0, \ldots, k' - 1\}$ we consider the hypergraph 
$G_{i_{1}, i_{2}, \ldots, i_{l}} = \cup_{S \subseteq \{1, 2, \ldots, k'\}: |S| = \frac{k}{l}} e_{i_{1}, i_{2}, \ldots, i_{l}}(S)$.
Let $G_{X_{1}, \ldots, X_{l}}$ be the hypergraph consisting of the union of all edges in $G_{i_{1}, i_{2}, \ldots, i_{l}}$ for every $i_{1}, i_{2}, \ldots, i_{l} \in \{0, \ldots, k' - 1\}$. Note that $G_{X_{1}, \ldots, X_{l}}$ has  $k'^{l} \cdot \dbinom{k'}{\frac{k}{l}}$ edges, as claimed.
It remains to show that every 2-coloring $c$ which has the same majority in $X_{1}, \ldots, X_{l}$ yields a monochromatic edge.
\begin{prop} \label{prop:existenceofappropriateshifting}
Let $s \in \{\text{red}, \text{blue}\}$ and let $c$ be a 2-coloring which has an $s$-majority in $X_{i}$ for every $i$, $i = 1, \ldots, l$. Then there are $i_{1}, i_{2}, \ldots, i_{l}$ such that for $\frac{k}{l}$ distinct $r$,
$x_{1, r + i_{1}}, x_{2, r + i_{2}}, \ldots, x_{l, r + i_{l}}$ all have color $s$ under $c$.
\end{prop}
\noindent
\emph{Proof:} Choose $i_{1}, i_{2}, \ldots, i_{l}$ uniformly at random from $\{0, 1, \ldots, k' - 1\}$. For every $r$ we let $Y_{r}$ be the indicator variable for the event that $x_{1, r + i_{1}}, x_{2, r + i_{2}}, \ldots, x_{l, r + i_{l}}$ all have color $s$ under $c$.
We have $\pr(Y_{r} = 1) \geq (\frac{1}{2})^{l}$.
So the expected value $E[\sum_{i = 1}^{k'} Y_{i}]$ is at least $k'(\frac{1}{2})^{l} = \frac{k}{l}$. Hence for some   
$i_{1}, i_{2}, \ldots, i_{l} \in \{0, 1, \ldots, k' - 1\}$, there are $\frac{k}{l}$ distinct $r$ where
$x_{1, r + i_{1}}, x_{2, r + i_{2}}, \ldots, x_{l, r + i_{l}}$ all have color $s$ under $c$. \hfill $\qed$

Let $r_{1}, r_{2}, \ldots, r_{\frac{k}{l}}$ be the distinct values for $r$ described in Proposition \ref{prop:existenceofappropriateshifting}. 
Let $S = \{r_{1}, r_{2}, \ldots, r_{\frac{k}{l}}\}$. Then $e_{i_{1}, i_{2}, \ldots, i_{l}}(S)$ is monochromatic under $c$. \hfill $\qed$

\emph{Proof of Proposition \ref{prop:evalformula}:} We use the following well-known fact. For every $r \leq n$,
\begin{equation} \label{eq:boundforbinomcoeff}
 \dbinom{n}{r} \leq \left( \frac{en}{r} \right)^{r}
\end{equation}
By \eqref{eq:boundforbinomcoeff},  $\dbinom{\frac{2^{l}}{l} k}{\frac{k}{l}} \leq \left(e2^{l} \right)^{k/l} = 2^{k} e^{k/l}$. Hence 
$m(k,l) \leq 2^{2l} \cdot 2^{l^{2}} \cdot k^{l} \cdot 2^{k} e^{\frac{k}{l}}$. Since $k^{\log k} = 2^{\log^{2} k}$ we get
$m(k,\log k) \leq  (2^{1 + o(1)})^{k}$. \hfill $\qed$

%\begin{thebibliography}{5}

%\end{thebibliography} 

%\bibliography{biblio}
% \bibliographystyle{plain}

\end{document}